%% file: maintext_arxiv.tex
\documentclass[a4paper,reprint,superscriptaddress,prb,aps]{revtex4-2}

\usepackage[english]{babel}
\usepackage[utf8]{inputenc}
\usepackage[T1]{fontenc}
\usepackage[version=4]{mhchem}

\usepackage{amssymb}
\usepackage{amsmath}
\usepackage{amsthm}
\usepackage{graphicx}
\usepackage{hyperref}
\usepackage{booktabs}
\usepackage{soul}
\usepackage[dvipsnames]{xcolor} 
\usepackage{tikz}
\usetikzlibrary{shapes.geometric, arrows}
\usepackage{bm}
\usepackage{listings}
\usepackage[per-mode=symbol,parse-numbers=false]{siunitx}

\makeatletter
\newcommand\footnoteref[1]{%
  \protected@xdef\@thefnmark{\ref{#1}}%
  \leavevmode
  \ifhmode\edef\@x@sf{\the\spacefactor}\nobreak\fi
  \@makefnmark
  \ifhmode\spacefactor\@x@sf\fi
  \relax}
\makeatother

\newcommand\abs[1]{\left\vert#1\right\vert}
\newcommand\dif{\mathrm{d}}
\newcommand\bra[1]{\left\langle#1\right\vert}
\newcommand\ket[1]{\left\vert#1\right\rangle}
\newcommand\braket[2]{\left\langle#1\middle\vert#2\right\rangle}

\newcommand\tdfrac[2]{\frac{\dif#1}{\dif#2}}
\newcommand\pfrac[2]{\frac{\partial#1}{\partial#2}}
\newcommand\kk{\mathbf{k}}

\colorlet{dkgreen}{green!60!black}

\newcommand{\psink}{\psi_{n\mathbf{k}}}
\newcommand{\unk}{u_{n\mathbf{k}}}

\newcommand{\rhopot}{V_\mathrm{xc}^\rho}
\newcommand{\taupot}{V_\mathrm{xc}^\tau}
\newcommand{\rhopotop}{\widehat{V}_\mathrm{xc}^\rho}
\newcommand{\taupotop}{\widehat{V}_\mathrm{xc}^\tau}
\def\batio3{\ce{BaTiO3}}

\newcommand{\repourl}{\url{https://github.com/epfl-matmat/supporting-mgga-addfpt}}

\begin{document}
\title{
Density functional perturbation theory of meta-generalized gradient approximations
using algorithmic differentiation
}

\author{Bruno Ploumhans}
\email{bruno.ploumhans@epfl.ch}
\author{Niklas Frederik Schmitz}
\author{Michael F. Herbst}
\email{michael.herbst@epfl.ch}
\affiliation{Mathematics for Materials Modelling (MatMat), Institute of Mathematics \& Institute of Materials,  École Polytechnique Fédérale de Lausanne, 1015 Lausanne, Switzerland}
\affiliation{National Centre for Computational Design and Discovery of Novel Materials (MARVEL), École Polytechnique Fédérale de
Lausanne, 1015 Lausanne, Switzerland}

\begin{abstract}
Density functional perturbation theory (DFPT) is an established framework for the computation of derivatives in plane-wave density functional theory (DFT).
We present an implementation of DFPT for exchange-correlation (XC) functionals $E_\mathrm{xc}(\rho,\tau)$ that incorporate an explicit dependence on both the density $\rho$ and the kinetic energy density $\tau$.
This covers the popular class of semilocal meta-generalized gradient approximations (meta-GGAs) as well as broader nonlocal parametrizations.
We sidestep the derivation of cumbersome XC second energy derivative expressions by recasting these derivatives as a Jacobian-vector product of the XC potentials,
which we evaluate with algorithmic differentiation (AD) techniques.
Integration with our previously developed AD-DFPT framework
\mbox{[N.~F.~Schmitz \textit{et~al.}, \textit{npj~Comput.~Mater.} \textbf{12}, 6 (2026)]}
provides access to derivatives of arbitrary ground state quantities
with respect to arbitrary perturbations.
We employ AD-DFPT to compute a range of response properties for ZnO and \batio3,
and find that the recent r2SCAN01 meta-GGA functional generally outperforms LDA and PBE.
Finally, we showcase the optimization of a neural-network meta-GGA
to self-consistently reproduce hybrid-DFT reference densities of bulk silicon,
using AD-DFPT gradients.
Overall, these results establish AD-DFPT as a versatile route for computing
DFT derivatives at the meta-GGA level, be they common response properties
or the unusual derivatives required for the gradient-based training of novel XC functionals.
\end{abstract}

\maketitle

\section{Introduction}
Simulations based on
density functional theory (DFT) are the most prevalent first-principles approach
to model the physical and chemical properties of materials.
Within the plane-wave DFT framework,
density functional perturbation theory (DFPT)~\cite{Gonze1997,Baroni2001}
is a method of choice for studying the response properties of materials
with respect to external perturbations.
As such it has been vital to the accurate prediction of a wide range of properties, including
phonon modes~\cite{baroni1987green,giannozzi1991}, dielectric constants~\cite{baroni1987elastic,gonze1992cg},
elastic constants~\cite{baroni1987elastic,hamann2005metrictensor},
piezoelectricity~\cite{deGironcoli1989piezoelectric,wu2005systematic},
nonlinear optical responses~\cite{veithen2005nonlinear},
effective masses~\cite{laflammejanssen2016},
electron-phonon coupling~\cite{giustino2017elph}
and flexoelectricity~\cite{royo2019flexoelectric}.
Compared to approaches employing finite differences,
DFPT is usually less sensitive to numerical noise, does not require step size tuning,
and can be more efficient~\cite{Baroni2001,Gonze1997}.
The main disadvantage of DFPT is the substantially increased implementation complexity
for each type of perturbation (e.g.~atomic displacement, strain)
and each differentiated quantity (e.g.~forces, stresses).

We recently proposed~\cite{ADpaper} making use of
forward-mode algorithmic differentiation (AD)~\cite{Griewank2008EDP,blondel_elements_2024} techniques
to replace the laborious manual implementation of perturbation-specific derivatives in DFPT 
and thus successfully tackle this biggest difficulty of DFPT.
In a nutshell, AD is an algorithmic technique to compute 
the derivative $f'(x)$ of a function $f(x)$
by working directly at the level of the computer code
implementing $f(x)$.
Ultimately this code performs a sequence of elementary operations
(floating-point operations, matrix operations, etc.).
AD differentiates each elementary operation using a database of known rules,
and propagates the derivative throughout the program using the chain rule.
Importantly, this computes the analytic derivative of $f(x)$ exactly and efficiently. 
In combination with a robust and efficient core solver for the DFPT problem itself,
our suggested AD-DFPT framework can thus in principle
compute the response of any quantity of interest to any perturbation~\cite{ADpaper}.

An emerging application of this framework is the computation of exchange-correlation (XC) parameter derivatives,
i.e.~the response of some property to changing a numerical parameter \textit{inside}
the XC functional~\cite{ADpaper}.
This relates to ongoing efforts towards machine-learned XC functionals,
where a crucial difficulty is the limited availability of training data
from higher level methods or experiments.
Although training XC functionals based on the energy remains the predominant approach%
~\cite{becke1997,mortensen2005beef,DM21,bystrom2024,Gao2025,luise2026skala,abdallah2026},
a number of works have incorporated additional reference data including
band gaps~\cite{Borlido2020,Kovacs2022,Bystrom2024gaps},
lattice constants~\cite{wellendorff2014mbeef,Kovacs2022},
as well as spatially resolved quantities such as
XC potentials~\cite{Riemelmoser2023,Kanungo2025},
correlation energy densities~\cite{Polak2025},
or molecular reference densities obtained by a
higher-level calculation~\cite{nagai2020,li2021ksregularizer,Kasim2021,dick2021,nagai2022,eberhard2026derivative}.
Such spatially resolved data
has been shown to greatly improve data efficiency, as more learning signal
can be extracted from each reference calculation.
To efficiently train an XC functional,
the gradient of a loss function including such quantities is then required.
While recent differentiable DFT codes~\cite{Zhang2022pyscf,Kasim2022,MCasares2024}
have made this feasible for Gaussian basis sets,
support has remained more scarce for plane-wave methods~\cite{ADpaper}.
This has so far hampered the exploration of, e.g.,
density-based learning strategies in the solid-state setting.
In this paper, we employ AD-DFPT
to compute the derivative of the self-consistent density wrt.~XC parameters,
thereby unlocking this promising path towards data-efficient XC functional learning
within plane-wave DFT.

Since solving a DFPT problem remains a core component of the AD-DFPT approach,
it is the generality of the DFPT solver itself, 
which determines which kind of DFT functionals can be treated within AD-DFPT.
In our first work~\cite{ADpaper} we employed a DFPT solver restricted
to DFT functionals based on the
generalized gradient approximation (GGA)~\cite{Perdew1996}.
In this work, we extend AD-DFPT towards meta-GGA functionals,
where the XC energy additionally
depends on the kinetic energy density (KED)~\cite{tpss2003,SCAN2015,Furness2020,LAK2024,Desmarais2025mscan}.
Meta-GGAs are currently under intense development and show promising results for a wide range of properties,
in particular for the prediction of structural properties~\cite{tran2016rungs},
phonon spectra~\cite{euchner2019phonons}, band gaps~\cite{LAK2024}, elastic properties~\cite{Haxhijaj2026}
and electron-phonon coupling~\cite{Wang2026}.
Notably,
their significant improvement in accuracy over GGAs is possible at only a marginally increased computational cost.
For similar reasons many recent
machine-learned XC functional parametrizations
incorporate an explicit dependence on the KED~\cite{wellendorff2014mbeef,nagai2020,dick2021,nagai2022,bystrom2024,luise2026skala}.
In spite of that, almost all widespread plane-wave DFT codes do not yet support DFPT for meta-GGA functionals.
This obstacle hinders the exploration of such meta-GGA functionals in materials simulations,
as accurately computing a wide range of response properties
is crucial to benchmarking and finding opportunities for further functional development.
The notable exception is a recent release of \textsc{CASTEP}~\cite{castep2005},
which after substantial implementation effort~\cite{durham2025thesis}
has extended its DFPT solver to meta-GGA functionals,
although only for a limited set of response properties.
In particular, we are unaware of any plane-wave-based code supporting the computation of
the XC parameter derivatives required to train KED-based machine-learned functionals.

On a technical level, a large obstacle to implementing DFPT
for functionals including the KED is that every XC functional parametrization
requires its own derivation of explicit expressions for the
second derivatives of the XC energy,
which entails substantial effort~\cite{refson2006castep_dfpt,sabatini2016vdwdfpt,durham2025thesis}.
Even when using a library such as \textsc{libxc}~\cite{lehtola2018libxc},
which directly provides many semilocal energy expressions and their derivatives,
or computing these pointwise derivatives with AD as suggested by \citeauthor{ekstrom2010}~\cite{ekstrom2010},
considerable implementation work is still required in the DFT code
to perform all the required contractions with perturbed densities.
\citeauthor{lehtola2026libxckernel} recognized this difficulty and recently proposed a new library to automate
the generation of this contraction layer~\cite{lehtola2026libxckernel}.

In this paper, we overcome this obstacle with an approach based on AD techniques:
we recast the computation of the second derivatives of the XC energy as
a Jacobian-vector product~(JVP) of the XC potentials,
which can be performed efficiently with forward-mode AD.
Of note, this approach is directly applicable to a general XC energy functional $E_\mathrm{xc}(\rho, \tau)$,
which may depend arbitrarily on $\rho$ and the KED $\tau$.
This covers the usual semilocal meta-GGAs,
but additionally includes density-based dispersion corrections~\cite{dion2004vdw,vv10,sabatini2013rvv10,ning2022r2scanrvv10},
and machine-learned XC parametrizations
built from nonlocal features~\cite{nagai2020,li2021ksregularizer,Riemelmoser2023,bystrom2024,Gao2025,luise2026skala,abdallah2026}.

Based on this development, we present an implementation of AD-DFPT
suitable for such general functionals $E_\mathrm{xc}(\rho, \tau)$.
At variance with the AD-DFPT framework of Ref.~\onlinecite{ADpaper} which makes use of AD \textit{around} a DFPT solver,
for the first time AD is additionally harnessed \textit{inside} the core DFPT solver
to evaluate the second derivatives of the XC energy.
As a demonstration, we compute for two relevant materials all second-order derivative tensors
for displacement, strain, and electric field perturbations.
We then showcase XC parameter derivatives by training a semilocal meta-GGA exchange functional against a few reference hybrid-DFT densities.

Throughout this work we treat the KED dependence according to the popular generalized Kohn--Sham formalism~\cite{seidl1996gks,yang2016gks,Perdew2017bandgaps}.
We refer the reader to the review of \citeauthor{kuemmel2008orbitaldft}~\cite{kuemmel2008orbitaldft} for the alternative optimized effective potential (OEP) method;
see also the letter of \citeauthor{nazarov2011mgga_kernel_oep}~\cite{nazarov2011mgga_kernel_oep} discussing the meta-GGA XC kernel in the OEP formalism.

The remainder of this paper is organized as follows.
Sec.~\ref{subsec:dft} reviews the self-consistent field (SCF) procedure for XC functionals involving
a dependence on the KED. %
The extension of AD-DFPT to this setting is then discussed across four sections.
Sec.~\ref{subsec:dfpt} presents the core DFPT procedure,
Sec.~\ref{subsec:potential_perturbations} develops our JVP-based approach
to obtaining second derivatives of the XC energy,
Sec.~\ref{subsec:ad_dfpt} describes how to compute response properties using AD-DFPT,
while Sec.~\ref{subsec:elfield} summarizes further specifics
of polarization responses and electric field perturbations.
Following an overview of our implementation and computational details~(Sec.~\ref{sec:impl})
we validate our AD-DFPT approach against finite differences in Sec.~\ref{subsec:elastic_diamond}.
Sec.~\ref{subsec:response_properties} illustrates the versatility of our AD-DFPT framework
by considering a wide range of response properties~%
(Born effective charges, relaxed-ion elastic constants, relaxed-ion piezoelectric tensor
and high-frequency and static dielectric tensor) using a variety of DFT functionals
up to the meta-GGA level.
Finally, Sec.~\ref{subsec:xc_learning} demonstrates the density-based learning
of a semilocal meta-GGA functional parametrized using a small neural network.

\section{Theory}
\label{sec:theory}

\subsection{Density functional theory}
\label{subsec:dft}

\begin{figure*}
    \centering
    \includegraphics[width=\textwidth]{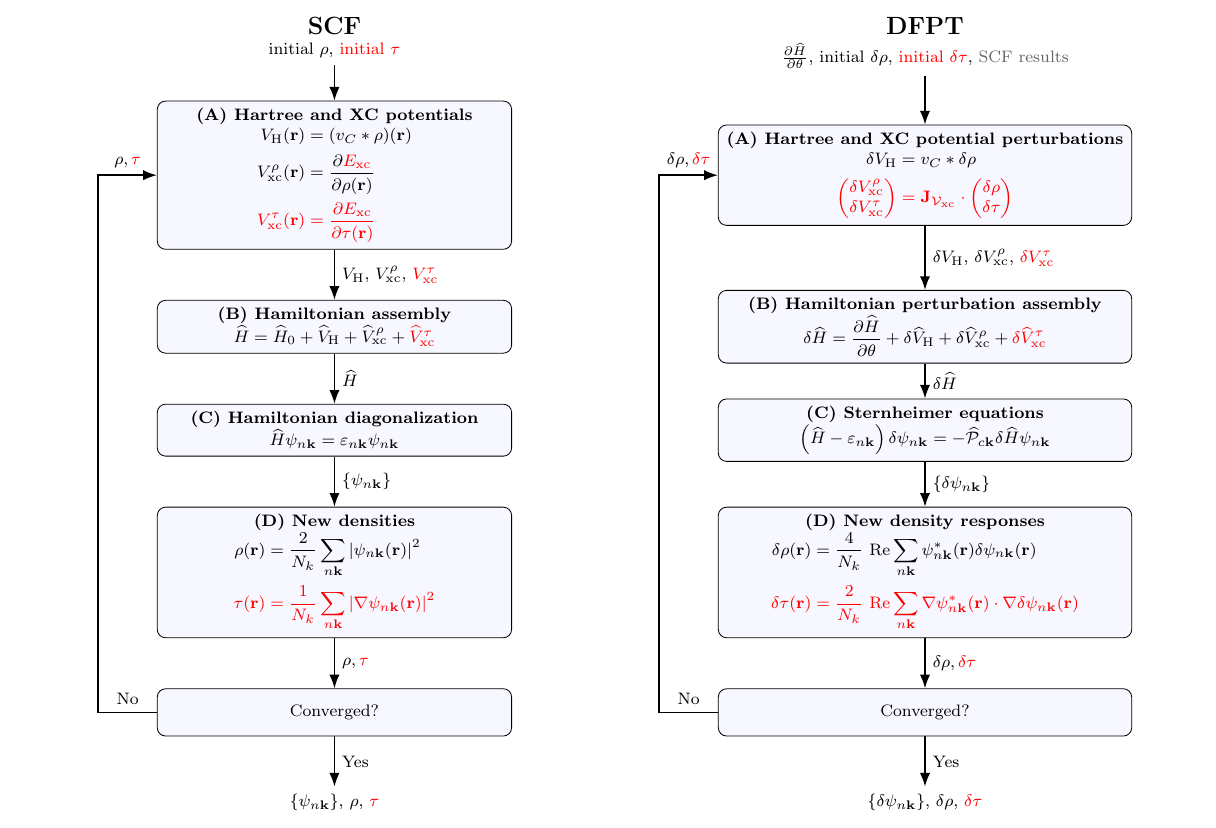}
    \caption{Ground state self-consistent field (SCF) iterations (left column) and density functional perturbation theory (DFPT) iterations (right column),
    for exchange-correlation (XC) functionals that depend on the kinetic energy density (KED).
    The two algorithms follow exactly the same structure, and each step of the DFPT iterations is the derivative of the corresponding SCF step.
    The additional terms that appear due to the KED are highlighted in red.}
    \label{fig:flowchart_scf_dfpt}
\end{figure*}

We briefly review DFT for
XC functionals including a dependence on the KED.
To simplify the presentation, we limit our discussion to insulating crystals
without spin polarization and employ atomic units throughout.
The additional treatment required for metallic systems or spin polarization is well established,
both for the ground-state and DFPT calculations~\cite{degironcoli1995dfptmetals,gonze2024variationaldfptmetals},
and extends easily to KED-based XC functionals.

Adopting a generalized Kohn--Sham formalism~\cite{seidl1996gks,yang2016gks,Perdew2017bandgaps},
the fundamental variables are the ground-state orbitals $\left\{\psink\right\}$.
They are obtained by minimization of the following total energy functional:
\begin{equation}
    \label{eq:total_energy}
    E[\left\{\psink\right\}] =
    \frac{2}{N_k} \sum_{n\mathbf{k}} \braket{\psink}{\widehat{H}_0\psink}
    + E_\mathrm{H}(\rho) + E_\mathrm{xc}(\rho, \tau).
\end{equation}

Here and in the rest of the paper, $n$ runs from 1 to half the number of electrons per unit cell of the crystal,
and the index $\mathbf{k}$ runs over the $N_k$ points of a Monkhorst--Pack discretization of the Brillouin zone.
In numerical implementations, crystal symmetries are exploited to reduce the number of independent $\mathbf{k}$-points,
but for simplicity we do not make this explicit and rather work with the reducible $\mathbf{k}$-point mesh.
Note that the density $\rho$ and the KED $\tau$ in Eq.~\eqref{eq:total_energy} should be understood as functions of the orbitals:
\begin{align}
    \label{eq:density_definition}
    \rho(\mathbf{r}) &= \frac{2}{N_k}\sum_{n\mathbf{k}} \abs{\psi_{n\mathbf{k}}(\mathbf{r})}^2, \\
    \label{eq:ked_definition}
    \tau(\mathbf{r}) &= \frac{1}{N_k} \sum_{n\mathbf{k}} \abs{\nabla \psi_{n\mathbf{k}}(\mathbf{r})}^2.
\end{align}

The three energy contributions in Eq.~\eqref{eq:total_energy} are defined as follows.
First, the non-interacting energy involving the non-interacting Hamiltonian $\widehat{H}_0$.
It is the sum of the kinetic energy
and the ionic potential energy (commonly expressed as a pseudopotential with local and nonlocal parts).
Second, the Hartree energy
\begin{equation}
    E_\mathrm{H}(\rho) = \frac12 \int_\Omega \dif\mathbf{r} \ \rho(\mathbf r) \left(v_C \ast \rho\right)(\mathbf r),
\end{equation}
where $\Omega$ is a unit cell, $v_C$ is the periodic Coulomb kernel
(including a compensating charge background),
and $\ast$ denotes a convolution.
Finally, the XC energy we take of the general form $E_\text{xc}(\rho, \tau)$,
i.e.~a general functional taking the functions $\rho$ and $\tau$ as input and returning a scalar energy.
We remark that this broad definition encapsulates
density-based dispersion corrections~\cite{dion2004vdw,vv10,sabatini2013rvv10,ning2022r2scanrvv10}
and recent nonlocal machine-learned parametrizations~\cite{nagai2020,li2021ksregularizer,bystrom2024,luise2026skala,abdallah2026}.
The common case of a semilocal meta-GGA is recovered by the usual definition
\begin{equation}
    \label{eq:semilocal_mgga}
    E_\mathrm{xc}(\rho, \tau) = \int_\Omega \dif\mathbf{r} \ e_\mathrm{xc}\left(\rho(\mathbf r), \nabla\rho(\textbf{r}), \nabla^2\rho(\textbf{r}), \tau(\mathbf r)\right),
\end{equation}
where $e_\mathrm{xc}$ is a local energy density,
and the derivatives $\nabla\rho$ and $\nabla^2\rho$ are defined
as smooth functions of $\rho$~\cite{whitebird1994}
in a plane-wave basis.

The goal of Kohn--Sham DFT is to find the orbitals that minimize the total energy.
The corresponding optimality conditions are derived by
differentiating the total energy Eq.~\eqref{eq:total_energy} wrt. trial orbitals.
Introducing the Lagrange multipliers $\left\{\varepsilon_{n\mathbf k}\right\}$ to enforce the orthonormality of the orbitals,
this leads to the celebrated Kohn--Sham equations:
\begin{align}
    \label{eq:eigenvalue_problem}
    \widehat{H}(\rho, \tau) \psink = \varepsilon_{n\mathbf k} \psink,
\end{align}
where the Hamiltonian is the operator $\widehat{H}(\rho,\tau)$
such that $\frac{N_k}{2} \pfrac{E}{\bra{\psink}} = \widehat{H}(\rho, \tau)\psink$.
These are nonlinear equations---the Hamiltonian $\widehat{H}(\rho, \tau)$ to be diagonalized depends on the eigenvectors through $\rho$ and $\tau$---%
which are usually solved using an iterative SCF procedure.

Figure~\ref{fig:flowchart_scf_dfpt} outlines this procedure in its first column.
The central step (C) is the solution of Eq.~\eqref{eq:eigenvalue_problem}
at a fixed $\widehat{H}$:
the equations are then linear eigenvalue problems,
which are solved by iterative diagonalization
for a new set of orbitals $\left\{\psink\right\}$.

Following the diagonalization,
a new density $\rho$ and KED $\tau$ is built
following Eqs.~\eqref{eq:density_definition} and \eqref{eq:ked_definition}.
This corresponds to step (D) in the figure.
Following that, a decision is made to either stop the SCF procedure,
or to continue iterating.
In our calculations, this decision is based on the
$L^2$-norm of the difference of the density across successive steps.

If the SCF procedure is continued, a new Hamiltonian is constructed by summing up the
term-wise derivatives of the total energy \eqref{eq:total_energy}:
\begin{align}
    \label{eq:jambon_assembly}
    \widehat{H}(\rho, \tau) = \widehat{H}_0 + \widehat{V}_\mathrm{H}(\rho) + \rhopotop(\rho, \tau) + \taupotop(\rho, \tau).
\end{align}
In particular, $\widehat{H}_0$ is the non-interacting Hamiltonian of Eq.~\eqref{eq:total_energy}
and $\widehat{V}_\mathrm{H}(\rho)$ is a local multiplication by the Hartree potential $V_\mathrm{H}(\mathbf r) = (v_C \ast \rho)(\mathbf r)$.
For the XC energy, the chain rule gives
\begin{align}
    \label{eq:xc_derivative}
    \pfrac{E_\mathrm{xc}}{\bra{\psink}} = \int_\Omega \dif\mathbf{r} \pfrac{E_\mathrm{xc}}{\rho(\mathbf r)} \pfrac{\rho(\mathbf r)}{\bra{\psink}}
    + \int_\Omega \dif\mathbf{r} \pfrac{E_\mathrm{xc}}{\tau(\mathbf r)} \pfrac{\tau(\mathbf r)}{\bra{\psink}}.
\end{align}
The first term leads to the usual local multiplication by the XC potential,
while the second term leads to an additional nonlocal operator that appears
due to the KED.
In this paper, we use the notation $\rhopot$ and $\taupot$
for the scalar fields $\rhopot(\mathbf r)=\pfrac{E_\mathrm{xc}}{\rho(\mathbf r)}$
and $\taupot(\mathbf r)=\pfrac{E_\mathrm{xc}}{\tau(\mathbf r)}$ respectively,
both of which we refer to as potentials.
The corresponding operators $\rhopotop$ and $\taupotop$
act as follows on trial orbitals~\cite{yao2017}:
\begin{align}
    \label{eq:rho_potential_application}
    \left(\rhopotop \psink\right)(\mathbf r) &= \rhopot(\mathbf{r}) \psink(\mathbf{r}), \\
    \label{eq:tau_potential_application}
    \left(\taupotop \psink\right)(\mathbf r) &= - \frac{1}{2} \nabla \cdot \left( \taupot(\mathbf{r}) \nabla \psink(\mathbf{r}) \right).
\end{align}

In Fig.~\ref{fig:flowchart_scf_dfpt}, the Hamiltonian construction is split in two steps.
First, the potentials $V_\mathrm{H}$, $\rhopot$, and $\taupot$ are constructed,
corresponding to step (A).
Then, the full Hamiltonian $\widehat{H}$ is assembled following Eq.~\eqref{eq:jambon_assembly},
corresponding to step (B).
The iterations then continue with the diagonalization of the newly assembled Hamiltonian
in step (C), which we discussed previously.
We have thus closed the loop,
concluding our description of the SCF iterations.

The main additions to the SCF procedure required to support 
XC functionals depending on the KED
are highlighted in red in Fig.~\ref{fig:flowchart_scf_dfpt}.
These are the terms involving the nonlocal potential $\taupotop$
as well as passing the KED $\tau$ along the iterations of the SCF, next to $\rho$.
We remark that for functionals of the density only,
various techniques are commonly used to accelerate the SCF
beyond the basic iterations presented here.
For an overview, we refer the reader to Refs.~\onlinecite{kresse1996efficient,herbst2020blackbox,chupin2021anderson}
and the references therein.
Not all these techniques are presently available for
functionals with a KED dependence
and their extension presents an interesting avenue for future work.

\subsection{Density functional perturbation theory}
\label{subsec:dfpt}

The total energy expression~\eqref{eq:total_energy}
implicitly depends on a number of modeling parameters
such as the atomic positions,
an applied strain, the strength of an applied electric field
or a parameter inside the XC functional;
let $\theta$ denote one such parameter.
As a result the orbitals minimizing
the total energy also acquire a dependence on $\theta$.
We write them as $\left\{\psink(\theta)\right\}$
and for simplicity assume them to be differentiable wrt.~$\theta$.
DFPT
is the framework that
enables the analytic computation
of the responses $\left\{\tdfrac{\psink}{\theta}\right\}$
of the orbitals to a perturbation of $\theta$,
starting from the results of a ground state computation.
In the rest of this section, we extend the usual DFPT formalism
taking the KED dependence of the XC functional into account.

In DFPT the orbital responses $\left\{\tdfrac{\psink}{\theta}\right\}$
are determined from the Hamiltonian perturbation
$\tdfrac{\widehat{H}}{\theta}$, which since we consider a single $\theta$
we write shortly as $\delta \widehat{H}$.
By the chain rule the total derivative of the Hamiltonian decomposes 
into a direct contribution (from its dependence on $\theta$)
plus additional terms induced by changes to $\rho$ and $\tau$
(which indirectly depend on $\theta$):
\begin{equation}
    \label{eq:full_deltaH}
    \delta \widehat{H} = \pfrac{\widehat{H}}{\theta} + \delta \widehat{V}_\mathrm{H} + \delta \rhopotop + \delta \taupotop,
\end{equation}
where $\delta \widehat{V}_\mathrm{H}$, $\delta \rhopotop$ and $\delta \taupotop$ are induced by
the following $\delta\rho$ and $\delta\tau$:
\begin{align}
    \label{eq:delta_n}
    \delta \rho(\mathbf{r}) &= \frac{4}{N_k}\ \mathrm{Re} \sum_{n\mathbf{k}} \psink^*(\mathbf{r}) \delta \psink(\mathbf r), \\
    \label{eq:delta_ked}
    \delta \tau(\mathbf{r}) &= \frac{2}{N_k}\ \mathrm{Re} \sum_{n\mathbf{k}} \nabla \psink^*(\mathbf{r}) \cdot \nabla \delta \psink(\mathbf r),
\end{align}
and where we also made use of the 
shorthand $\delta\psink = \tdfrac{\psink}{\theta}$.
Thus,
the Hamiltonian perturbation $\delta\widehat{H}$ must be evaluated from the density responses $\delta\rho$ and $\delta\tau$,
which depend on the orbital responses,
thus are themselves ultimately again computed from the Hamiltonian perturbation.
As a result, DFPT takes the form of a self-consistent cycle,
see the right hand side of Fig.~\ref{fig:flowchart_scf_dfpt}.
Notably, the DFPT cycle shows a close correspondence to the
SCF loop of the ground state computation:
each step of DFPT (right column) is obtained by taking
the derivative of a respective SCF step on the left.

At the core of DFPT is step (C) of Fig.~\ref{fig:flowchart_scf_dfpt},
the solution of the Sternheimer equations at a fixed $\delta\widehat{H}$.
These equations are obtained by formally differentiating the Kohn--Sham equations~\cite{Baroni2001},
yielding one linear system for each $n$ and $\mathbf{k}$,
\begin{align}
    \label{eq:projected_sternheimer}
    \left(\widehat{H} - \varepsilon_{n\mathbf{k}}\right) \delta\psink
    = - \widehat{\mathcal P}_{c\mathbf{k}} \delta \widehat{H} \psink,
\end{align}
which is solved for the new orbital responses $\left\{\delta\psink\right\}$.
Here, $\widehat{\mathcal P}_{c\mathbf{k}} = 1 - \sum_n \ket{\psink} \bra{\psink}$
is the projector onto the conduction bands.
For insulating crystals, the equations
are solved under the orthogonality condition
$\braket{\psink}{\delta\psi_{m\mathbf{k}}} = 0$
for all $m,n,\mathbf{k}$.
For metals, solving Eq.~\eqref{eq:projected_sternheimer}
only yields the conduction contribution to the orbital responses,
and an additional valence contribution
is computed with an explicit sum-over-states formula,
see Ref.~\onlinecite{ResponseCalculations}.

Continuing with the DFPT procedure of Fig.~\ref{fig:flowchart_scf_dfpt},
the density and KED responses $\delta\rho$ and $\delta\tau$
are evaluated in step (D)
from the newly obtained orbital responses $\left\{\delta\psink\right\}$,
following Eqs.~\eqref{eq:delta_n} and \eqref{eq:delta_ked}.
Convergence is then determined from the change in $\delta\rho$
between two successive DFPT cycles.

If the DFPT procedure is continued,
the potential perturbations $\delta V_\mathrm{H}$, $\delta\rhopot$, and $\delta\taupot$
are computed,
corresponding to step (A) in Fig.~\ref{fig:flowchart_scf_dfpt}.
While the Hartree potential perturbation is standard and obtained by
differentiating $V_\mathrm{H}$ wrt. $\rho$,
\begin{equation}
    \delta V_\mathrm{H}(\mathbf r) = (v_C \ast \delta \rho)(\mathbf r),
\end{equation}
we employ AD to compute the XC potential perturbations,
as detailed in Sec.~\ref{subsec:potential_perturbations}.

From the potential perturbations
(as well as $\pfrac{\widehat{H}}{\theta}$), the
full Hamiltonian perturbation $\delta\widehat{H}$ is assembled in step (B),
following Eq.~\eqref{eq:full_deltaH}.
From $\delta\widehat{H}$,
the right-hand side of the Sternheimer equations~\eqref{eq:projected_sternheimer}
can be computed.
Note that the potential perturbations are also
applied as operators, namely the
local operators $\delta \widehat{V}_\mathrm{H}$
and $\delta \rhopotop$ according to Eq.~\eqref{eq:rho_potential_application}
and the nonlocal operator $\delta \taupotop$
according to Eq.~\eqref{eq:tau_potential_application}.
Finally, the Sternheimer equations are solved in step (C), as discussed above.
This closes the loop of the DFPT procedure.

The main structural changes introduced by the KED and its response are
highlighted in red in Fig.~\ref{fig:flowchart_scf_dfpt}.
These are the additional nonlocal XC potential perturbation $\delta\taupotop$
and the extra $\delta\tau$,
which is carried across DFPT iterations alongside $\delta\rho$.
In particular, the solution of the Sternheimer equations is agnostic to
the precise form of the Hamiltonian perturbation $\delta\widehat{H}$.
Existing Sternheimer solvers can therefore be employed
without modification
as long as the additional nonlocal Hamiltonian terms
$\taupotop$ and $\delta\taupotop$ are taken into account.

As with the ground state SCF procedure,
various techniques can be used to accelerate DFPT iterations.
In our implementation, the DFPT fixed-point problem is recast as a linear system in the unknowns $\delta\rho$ and $\delta\tau$,
and solved using preconditioned inexact Krylov methods as described in
Ref.~\onlinecite{InexactKrylovResponse}.

\subsection{XC potential perturbations using algorithmic differentiation}
\label{subsec:potential_perturbations}

This section covers the computation of
the XC potential perturbations $\delta\rhopot$ and $\delta\taupot$
in step (A) of the DFPT cycle in Fig.~\ref{fig:flowchart_scf_dfpt}.

Since $\rhopot$ and $\taupot$ each depend on both $\rho$ and $\tau$,
a general expression for the induced XC potential perturbations is
\begin{equation}
    \label{eq:mgga_kernel}
    \begin{pmatrix} \delta \rhopot(\mathbf r) \\ \delta \taupot(\mathbf r) \end{pmatrix}
    = \int_\Omega \dif\mathbf{r'} \begin{pmatrix}
        \pfrac{\rhopot(\mathbf r)}{\rho(\mathbf r')} & \pfrac{\rhopot(\mathbf r)}{\tau(\mathbf r')} \\
        \pfrac{\taupot(\mathbf r)}{\rho(\mathbf r')} & \pfrac{\taupot(\mathbf r)}{\tau(\mathbf r')} \\
    \end{pmatrix} \cdot
    \begin{pmatrix} \delta \rho(\mathbf r') \\ \delta \tau(\mathbf r') \end{pmatrix}.
\end{equation}
Implementing this equation is one of the key
challenges in extending an existing DFPT code to
support XC functionals depending on the KED.
Indeed, the kernel matrix in Eq.~\eqref{eq:mgga_kernel}
must be derived and implemented for each XC functional form,
which can lead to cumbersome expressions and requires substantial effort.
See, for example, the detailed discussion of this kernel expression
for GGA functionals~\cite{refson2006castep_dfpt},
van der Waals functionals~\cite{sabatini2016vdwdfpt},
and more recently semilocal meta-GGAs~\cite{durham2025thesis}.

In this work we compute the XC potential perturbations
automatically using AD techniques.
Denote by $\mathcal{V}_\mathrm{xc}$ the XC densities-to-potentials map
\begin{equation}
    \mathcal V_\mathrm{xc}: (\rho, \tau) \mapsto (\rhopot, \taupot).
\end{equation}
Such a function, or an equivalent thereof, is already used in existing DFT codes
to evaluate the XC potentials in step (A) of the ground state SCF cycle.
The key observation is that the response of the XC potentials in Eq.~\eqref{eq:mgga_kernel} is the product of the Jacobian $\mathbf{J}_{\mathcal V_\mathrm{xc}}$ of $\mathcal V_\mathrm{xc}$
with the vector $(\delta\rho, \delta\tau)$, namely
\begin{equation}
    \label{eq:mgga_kernel_jvp}
    \begin{pmatrix} \delta \rhopot \\ \delta \taupot \end{pmatrix}
    = \mathbf{J}_{\mathcal V_\mathrm{xc}} \cdot \begin{pmatrix} \delta \rho \\ \delta \tau \end{pmatrix},
\end{equation}
where the Jacobian is evaluated at $(\rho,\tau)$.

Computing a Jacobian-vector product (JVP) is a standard primitive in differentiable programming,
and it can be evaluated automatically and efficiently using forward-mode AD.
In our implementation, we fix $\rho$, $\tau$, $\delta\rho$, and $\delta\tau$, and
introduce the following function
\begin{equation}
    \label{eq:f_xc}
    f(\varepsilon) = \mathcal V_\mathrm{xc}(\rho + \varepsilon \cdot \delta\rho, \tau + \varepsilon \cdot \delta\tau),
\end{equation}
where $\varepsilon$ is a real number.
We then apply forward-mode AD to compute $f'(0)$.
By the chain rule, $f'(0) = (\delta \rhopot, \delta \taupot)$,
i.e.~the derivative
yields exactly the JVP \eqref{eq:mgga_kernel_jvp} and thus the XC potential perturbations.
The cost of evaluating the XC potential perturbations in this manner
is, up to a small constant factor (roughly two~\cite[Chapter 3]{Griewank2008EDP}),
the same as that of evaluating
the XC potentials themselves,
and thus negligible compared to the entire cost of the DFPT calculation.

While this approach thus enables automatic
computation of the XC potential perturbations with full generality,
it requires an implementation of $\mathcal V_\mathrm{xc}$ that
can be treated with AD tools.
For XC functionals based on common machine-learning frameworks~\cite{pytorch,jax2018github}
this is usually a given.
In fact, typically the XC potentials $\mathcal V_\mathrm{xc}$
themselves are obtained by differentiating the XC energy expression.
The perturbations of Eq.~\eqref{eq:mgga_kernel_jvp}%
---i.e. a Hessian-vector product of the energy---%
thus only requires calling AD a second time~\cite{paerlmutter1994hvp}.

In contrast, for traditional semilocal meta-GGA functionals,
the evaluation of $\mathcal V_\mathrm{xc}$ typically involves
a call to an external library such as \textsc{libxc}~\cite{lehtola2018libxc}
to obtain the values of $\rhopot$ and $\taupot$ on a grid.
Such external calls cannot be differentiated out of the box,
but one can supply custom AD rules,
which effectively teach the AD system how to
request the appropriate second derivatives from \textsc{libxc}.
As a result, the AD system can still automatically differentiate
through \emph{all other} operations of $\mathcal V_\mathrm{xc}$%
---including fast Fourier transforms and contractions with the density.

\subsection{The AD-DFPT framework for response property computations}
\label{subsec:ad_dfpt}
From the SCF orbitals $\left\{\psink(\theta)\right\}$
a wide range of DFT ground state quantities of interest~(QoIs)
\begin{equation}
    \label{eq:a_definition}
    A (\{\psink(\theta)\}, \theta)
\end{equation}
are accessible, such as the forces, the stresses
or any other Hellmann--Feynman derivative of the total energy.
Recall that $\theta$ denotes an arbitrary
modeling parameter of the DFT calculation.
By choosing a suitable $A$ and $\theta$, many material properties
can in fact be expressed as a derivative $\tdfrac{A}{\theta}$.
The cases we consider in this work are 
summarized in Tab.~\ref{tab:response_properties}.
We will additionally consider in Sec.~\ref{subsec:xc_learning}
the training of a new XC functional where $A$ will be a loss against
a hybrid-DFT target and $\theta$ the parameters of a neural network.

\begin{table}
    \centering
    \begin{tabular}{l@{\hskip 0.5cm}l}
        \toprule
        Response tensor & Definition \\
        \midrule
        Force-constant matrix & $K_{s\alpha,t\beta} = -\tdfrac{F_{t\beta}}{u_{s\alpha}}$ \\
        Force-response internal strain & $\Lambda_{s\alpha j} = \tdfrac{F_{s\alpha}}{\eta_j}$ or $-\Omega\tdfrac{\sigma_j}{u_{s\alpha}}$ \\
        Born effective charges & $Z_{s\alpha\beta}^\ast = \tdfrac{F_{s\alpha}}{\mathcal E_\beta}$ or $\Omega \tdfrac{P_\beta}{u_{s\alpha}}$ \\
        Cl.-ion elastic tensor & $\bar{C}_{jk} = \tdfrac{\sigma_k}{\eta_j}$ \\
        Cl.-ion piezoelectric tensor & $\bar{e}_{\alpha j} = -\tdfrac{\sigma_j}{\mathcal E_\alpha}$ or $\tdfrac{P_\alpha}{\eta_j}$ \\
        Cl.-ion dielectric susceptibility & $\bar{\chi}_{\alpha\beta} = \tdfrac{P_\beta}{\mathcal E_\alpha}$ \\
        \bottomrule
    \end{tabular}
    \caption{Definition of the elementary response tensors.
    $F_{s\alpha}$, $\sigma_j$, $P_\alpha$ are the forces, stresses, and polarization,
    while
    $u_{s\alpha}$, $\eta_j$, $\mathcal E_\alpha$ are the atomic positions, strain, and electric field, respectively.
    Our index convention is that $s$ and $t$ refer to atoms, $\alpha$ and $\beta$ to the three Cartesian directions
    and $j$ and $k$ to the six strain components in Voigt notation.
    $\Omega$ is the volume of the unit cell. The last three tensors are clamped-ion (cl.-ion) quantities.
    In this work we compute the proper piezoelectric tensor, see the discussion in Refs.~\onlinecite{wu2005systematic,Vanderbilt2000}.
    }
    \label{tab:response_properties}
\end{table}

Leaving the additional technicalities for computing polarization responses
and electric field derivatives to Sec.~\ref{subsec:elfield},
such derivatives can indeed be analytically computed with AD-DFPT.
In fact, as demonstrated previously~\cite{ADpaper} defining the expression for $A(\{\psink(\theta)\}, \theta)$
and subsequently requesting a derivative $\tdfrac{A}{\theta}$ is sufficient
to implement a new DFT derivative or a new DFT property.
The main achievement of this work consists in extending this approach
to XC functionals depending on the KED.

In a nutshell, upon requesting $\tdfrac{A}{\theta}$, our AD-DFPT framework
will compute this derivative following the following three steps.
First, the algorithm computes 
$\pfrac{\widehat{H}}{\theta}$ by literally differentiating the routine building the
Hamiltonian with respect to the considered parameter $\theta$ with AD.
Notably, this conveniently treats all terms of the Hamiltonian automatically.
For example, in the case of strain perturbations,
AD will correctly sum the contributions from all the Hamiltonian terms~\cite{hamann2005metrictensor},
which in the case of meta-GGA functionals includes correctly summing the 
contributions from both $\rhopotop$ and $\taupotop$.
Next, it invokes our DFPT solver newly extended to KED-based functionals (Sec.~\ref{subsec:dfpt})
to compute $\left\{\delta\psink\right\}$.
Finally, the total derivative is assembled by the chain rule
\begin{multline}
    \label{eq:e2e_derivative}
    {\tdfrac{A}{\theta}
    = \pfrac{A}{\theta} + \sum_{n \mathbf{k}} \int_\Omega \dif \mathbf{r} \ \pfrac{A}{\psink(\mathbf r)} \delta\psink(\mathbf r)} \\
    + \sum_{n \mathbf{k}} \int_\Omega \dif \mathbf{r} \ \pfrac{A}{\psink^*(\mathbf r)} \delta\psink^*(\mathbf r).
\end{multline}
This quantity can again be simply evaluated as a JVP on $A$,
at $(\left\{\psink\right\}, \theta)$ with the perturbation $(\left\{\delta\psink\right\}, 1)$.

\subsection{Polarization responses and electric field perturbations}
\label{subsec:elfield}
DFPT is commonly used to compute
the response of the electric polarization to perturbations,
as well as the related response to electric field perturbations.
In the notation of the previous sections, this corresponds to computing derivatives
$\tdfrac{A}{\theta}$ where either $A$ is the electric polarization, or $\theta$ is
the strength of an applied electric field, or both.
For extended systems it is well known that obtaining such responses
in a physically meaningful manner requires a
suitable definition of the polarization, respectively a suitable definition
of the energy functional including an electric field strength.
We obtain those following the 
Berry phase theory of polarization~\cite{KingSmith1993polarization,resta1994,Nunes2001pead,souza2002finitefields}.
This effectively leads to minor modifications of the
overall DFPT workflow and the form of Sternheimer equations that need solving.
As detailed below these modifications %
are mostly restricted to inserting appropriate pre-processing or post-processing
step to the DFPT procedure.

\subsubsection{Polarization responses}
\newcommand{\param}{p}
In the Berry phase theory of polarization
the cell-periodic parts $\left\{\unk\right\}$
of the Bloch orbitals, i.e.~$\psink(\mathbf r) = \unk(\mathbf r) e^{i\kk\cdot\mathbf{r}}$,
are the fundamental variables to consider.
In particular, the response of the electronic polarization $\mathbf{P}^\mathrm{el}$
to a perturbation of $\theta$ is evaluated as~\cite[Eq. (60)]{resta1994}
\begin{equation}
    \label{eq:p_el_derivative}
    \tdfrac{P^\mathrm{el}_\alpha}{\theta} = \frac{4}{\Omega N_k} \mathrm{Im} \sum_{n\kk} \braket{\tdfrac{\unk}{\theta}}{\tdfrac{\unk}{k_\alpha}},
\end{equation}
where $\alpha$ denotes a Cartesian direction.
We thus require orbital derivatives of the form $\tdfrac{\unk}{\param}$,
where $\param=\theta$ or $\param=k_\alpha$.
Introducing $\widehat{H}_\kk = e^{-i\kk\cdot\widehat{\mathbf{r}}} \widehat{H} e^{i\kk\cdot\widehat{\mathbf{r}}}$
we can rewrite the Sternheimer equations~\eqref{eq:projected_sternheimer} as
\begin{equation}
    \label{eq:sternheimer_k}
    (\widehat{H}_\kk - \varepsilon_{n\kk}) \tdfrac{\unk}{\param}
    = - \widehat{\mathcal P}_{c\kk} \tdfrac{\widehat{H}_\kk}{\param} \unk,
\end{equation}
with $\widehat{\mathcal P}_{c\kk}$ redefined as $1 - \sum_n \ket{\unk} \bra{\unk}$.
To compute the polarization response \eqref{eq:p_el_derivative}
this set of equations has to be solved for both $\param=\theta$ and $\param=k_\alpha$
(and each time for all bands $n$ and all $\kk$).

Since the mathematical structure of Eq.~\eqref{eq:sternheimer_k}
remains closely related to Eq.~\eqref{eq:projected_sternheimer}---with effectively a modified right-hand side
and the unknowns $\tdfrac{\unk}{\param}$ instead of $\tdfrac{\psi_{n\kk}}{\param}$---%
the same algorithmic procedure already outlined in Sec.~\ref{subsec:dfpt} can be employed.
Notably, the derivatives $\tdfrac{\widehat{H}_\kk}{\theta}$ in general feature induced terms
\begin{equation}
    \label{eqn:Hinduced_k}
    \delta \widehat{H}_{\mathbf{k},\mathrm{ind}} = e^{-i\mathbf{k}\cdot\widehat{\mathbf{r}}} \left(\delta\widehat{V}_\mathrm{H} + \delta\rhopotop + \delta\taupotop \right) e^{i\mathbf{k}\cdot\widehat{\mathbf{r}}},
\end{equation}
which depend on the responses $\delta \rho$ and $\delta \tau$.
For computing $\tdfrac{\unk}{\theta}$ we thus require a self-consistency cycle
and multiple Sternheimer solves.
In contrast, the derivative of the density and KED wrt.~$\kk$ vanishes,
such that $\tdfrac{\widehat{H}_\kk}{k_\alpha}$ has a zero induced
Hamiltonian $\delta \widehat{H}_{\mathbf{k},\mathrm{ind}}$
and a single Sternheimer solve is sufficient to obtain $\tdfrac{\unk}{k_\alpha}$
for each $n$, $\mathbf{k}$, $\alpha$.

On top of our standard AD-DFPT
implementation outlined in Secs.~\ref{subsec:dfpt} to~\ref{subsec:ad_dfpt},
only two new ingredients are thus needed,
namely (i) the derivatives
$\tdfrac{\widehat{H}_\kk}{\theta}$ and $\tdfrac{\widehat{H}_\kk}{k_\alpha}$
and (ii) an additional inner product \eqref{eq:p_el_derivative}.
Both require only minimal manual effort as we compute the Hamiltonian derivatives in (i) again
by a single function call to invoke the AD system. %

Of note, this remains entirely general to the type of XC functionals employed
in the DFT computation---provided the underlying DFPT solver can treat respective functionals.
In particular,
for functionals with a KED dependence this notably automatically takes care of including all contributions
from the kinetic energy operator, the nonlocal pseudopotential,
and the nonlocal potential $\taupotop$.

\subsubsection{Electric field perturbations}
Consider a material under zero macroscopic electric field
and apply an infinitesimal electric field of strength $\mathcal E_\alpha$
along Cartesian direction $\alpha$. Our goal is to compute
the response $\tdfrac{A}{\mathcal E_\alpha}$ for a generic QoI
$A\left(\left\{\unk(\mathcal E_\alpha)\right\},E_\alpha\right)$.
Within our AD-DFPT framework of Sec.~\ref{subsec:ad_dfpt} we compute this response by the chain rule
\begin{multline}
    \label{eq:e2e_elfield}
    {\tdfrac{A}{\mathcal E_\alpha}
    = \pfrac{A}{\mathcal E_\alpha} + \sum_{n \mathbf{k}} \int_\Omega \dif \mathbf{r} \ \pfrac{A}{\unk(\mathbf r)} \tdfrac{\unk(\mathbf r)}{\mathcal E_\alpha}} \\
     + \sum_{n \mathbf{k}} \int_\Omega \dif \mathbf{r} \ \pfrac{A}{\unk^*(\mathbf r)} \tdfrac{\unk^*(\mathbf r)}{\mathcal E_\alpha},
\end{multline}
i.e.~by asking the AD system to perform a
JVP on $A$, this time evaluated at $\left(\{\unk\}, 0\right)$
with perturbation $\big(\big\{\tdfrac{\unk}{\mathcal E_\alpha}\big\}, 1\big)$.

Following DFPT the required orbital responses %
are formally obtained by solving the Sternheimer equations
\begin{equation}
    (\widehat{H}_\mathbf{k} - \varepsilon_{n\mathbf{k}}) \tdfrac{\unk}{\mathcal E_\alpha}
    = - \widehat{\mathcal P}_{c\mathbf{k}} \left(\pfrac{\widehat{H}_{\mathbf{k}}}{\mathcal E_\alpha} \unk + \delta \widehat{H}_{\mathbf{k},\mathrm{ind}} \unk\right),
\end{equation}
where we explicitly separated the Hamiltonian derivative
into the direct and induced term \eqref{eqn:Hinduced_k}.
For an energy functional including an electric field term,
the Hamiltonian operator cannot be written at a nonzero electric field,
such that $\pfrac{\widehat{H}_{\mathbf{k}}}{\mathcal E_\alpha}$ is only a conceptual expression~\cite{Nunes2001pead}.
According to the arguments of \citeauthor{Nunes2001pead}~\cite{Nunes2001pead} we instead
replace $\pfrac{\widehat{H}_{\mathbf{k}}}{\mathcal E_\alpha} \unk$
by $i\tdfrac{\unk}{k_\alpha}$ when solving for the responses $\big\{\tdfrac{\unk}{\mathcal E_\alpha}\big\}$.

From the implementation point of view we therefore first evaluate
the responses $\big\{\tdfrac{\unk}{k_\alpha}\big\}$
as a pre-processing step, i.e.~by solving Eq.~\eqref{eq:sternheimer_k} non-self-consistently
for $\param=k_\alpha$.
Next we supply the right-hand sides
$\big\{i \tdfrac{\unk}{k_\alpha}\big\}$%
to a second self-consistent DFPT problem solved as described in Sec.~\ref{subsec:dfpt}.
The final ingredient is the JVP invocation within the AD system
to compute Eq.~\eqref{eq:e2e_elfield}.
In the case where $A$ is the electronic polarization,
Eq.~\eqref{eq:p_el_derivative} is used instead.

\section{Implementation and computational details}
\label{sec:impl}
Unless otherwise noted,
we perform all the calculations with a modified version of the Density-Functional Toolkit (\textsc{DFTK}) v0.7.25~\cite{DFTKpaper}.
This version of \textsc{DFTK} already supported the SCF procedure described in Sec.~\ref{subsec:dft}
for meta-GGA XC functionals,
accelerated by mixing the density across SCF steps.
We extended its existing DFPT solver implementation~\cite{ResponseCalculations,InexactKrylovResponse}
to support meta-GGA functionals, following Sec.~\ref{subsec:dfpt} to~\ref{subsec:elfield}.
An integration within \textsc{DFTK} is planned for a future release.
The \textsc{ForwardDiff} v1.4.1~\cite{Revels2016forwarddiff} package is used for all AD operations
and we use the \textsc{libxc}~\cite{lehtola2018libxc} v7.0.0 implementation of
the Perdew--Wang local density approximation (LDA)~\cite{perdew1992lda},
the Perdew--Burke--Ernzerhof (PBE) GGA~\cite{Perdew1996} and
the r2SCAN01 meta-GGA~\cite{Holzwarth2022r2scan01}.
The reference implementation, employed crystal structures as well as full computational
details can be found in the supporting code repository at~\repourl~\cite{supporting_repo}.

For the meta-GGA property computations,
we found significant discrepancies when combining GGA pseudopotentials
with meta-GGA XC functionals, in agreement with previous work~\cite{yao2017}.
Hence, we use norm-conserving pseudopotentials generated by the \textsc{METAPSP}~\cite{metapsp} code,
a development version of \textsc{ONCVPSP}~\cite{Hamann2013oncv} capable of generating meta-GGA pseudopotentials.
We used the alpha version 1.0.1 05/16/2024 of \textsc{METAPSP},
with minor bug-fixing patches, see details in the supporting code repository.
The only meta-GGA XC functional supported by this version is the r2SCAN01~\cite{Holzwarth2022r2scan01} functional,
a version of r2SCAN~\cite{Furness2020} with the regularization parameter $\eta$ increased from $0.001$ (in r2SCAN)
to $0.01$ (in r2SCAN01).
We thus use r2SCAN01 for most of our experiments to ensure consistency with the pseudopotentials.
We validated our pseudopotentials against r2SCAN01 all-electron (AE) reference data~\cite{mgga_ae_data}
following the protocol of~\citeauthor{Bosoni2023}~\cite{Bosoni2023},
see the Supplementary Materal (SM) for more details.

\subsection{Diamond} 
For the test detailed in~Sec.~\ref{subsec:elastic_diamond}
we use a C pseudopotential generated by \textsc{METAPSP} for r2SCAN01.
The PBE parameters from the PseudoDojo pseudopotential~\cite{van_setten_pseudodojo_2018} were used as a starting point,
and then modified to improve agreement with the AE reference~(see supplementary section~\ref{supp:pseudos}).
We use an energy cutoff of 55~Ha with the cutoff smearing scheme of Ref.~\onlinecite{BlowupCHV},
an $8\times8\times8$ grid of \mbox{$\mathbf{k}$-points}
and Fermi--Dirac smearing with a temperature of 1~mHa.
The SCF and DFPT computations are stopped when a very tight $10^{-12}$
threshold is reached in the density norm.
The structure was obtained by relaxing a bulk diamond structure generated with
Atomic Simulation Environment~\cite{ASE} 
until the largest force was smaller than \SI{10^{-4}}{\eV\per\angstrom}
and the largest stress component was smaller than \SI{0.1}{\kilo\bar}.

\subsection{ZnO and \batio3}
For our calculations in Sec.~\ref{subsec:response_properties}
initial structures were obtained from
the Materials Cloud Three-Dimensional Structure Database~\cite{MC3D},
identifiers \texttt{mc3d-31594} and \texttt{mc3d-84978}.
These were subsequently relaxed until the largest force was smaller than \SI{10^{-4}}{\eV\per\angstrom}
and the largest stress component was smaller than \SI{0.1}{\kilo\bar}.
For the LDA and PBE calculations, we use the v0.4.1 standard PseudoDojo pseudopotentials~\cite{van_setten_pseudodojo_2018},
with a \textit{high} energy cutoff of 48~Ha.
For the r2SCAN01 calculations, we use Ba, O, Ti, and Zn pseudopotentials
generated by \textsc{METAPSP},
with an energy cutoff of 60~Ha;
the PBE parameters from the PseudoDojo pseudopotentials were
used as is, and found to provide sufficient agreement with the AE reference,~see supplementary section~\ref{supp:pseudos}.
Calculations on ZnO employed a $12\times12\times8$ $\mathbf{k}$-point grid,
while those on \batio3 employed a $8\times8\times8$ grid.
In all cases we use Fermi--Dirac smearing with a temperature of 1~mHa,
the energy cutoff smearing scheme of Ref.~\onlinecite{BlowupCHV}
and stop both SCF and DFPT computations when the density norm difference is below $10^{-8}$.

\subsection{XC functional optimization}
\label{subsec:mlxcdetails}

For the application in Sec.~\ref{subsec:xc_learning},
the HSE06 (Heyd--Scuseria--Ernzerhof)~\cite{hse03,hse06} hybrid reference calculations
were performed with \textsc{Quantum ESPRESSO}~\cite{qe2009,qe2017} v7.5;
all other calculations
are performed with DFTK.
All calculations use the following identical numerical settings.
We use the v1.2 Schlipf--Gygi Si pseudopotential~\cite{sg15}
as it is built without a nonlinear core correction,
with an energy cutoff of 25~Ha.
We use an $8\times8\times8$ \mbox{$\mathbf k$-grid} combined with 1 mHa Gaussian smearing,
with the exact-exchange $\mathbf q$-grid of HSE06 taken equal to the $\mathbf k$-grid.
The $\mathbf k$-grid is not fully converged for the metallic
$\beta$-tin phase, but the comparison is consistent between functionals.

The exchange enhancement factor $F_\theta(\rho, |\nabla\rho|, \tau)$
is parametrized pointwise using
the bounded variables
\begin{equation}
    u = \frac{s^2}{1+s^2}, \qquad
    w = \frac{\tau_\mathrm{unif} - \tau}{\tau_\mathrm{unif} + \tau},
\end{equation}
where $s = \frac{|\nabla\rho|}{2(3\pi^2)^{1/3}\rho^{4/3}}$ is the reduced density gradient,
and $\tau_\mathrm{unif} = \tfrac{3}{10}(3\pi^2)^{2/3}\rho^{5/3}$ is the KED of the uniform electron gas (UEG).
Both $u$ and $w$ vanish in the UEG limit
and were first introduced in Refs.~\onlinecite{becke1997} and \onlinecite{becke2000} respectively.
We then define the enhancement factor as
\begin{equation}
    F_\theta(u, w) = 1 + N_\theta(u, w) - N_\theta(0, 0),
\end{equation}
with $N_\theta$ a multilayer perceptron with one hidden layer of width $8$ and
softplus activation, giving $33$ learnable parameters. Subtracting $N_\theta(0,0)$
ensures that the UEG limit $F_\theta(0, 0) = 1$ is satisfied for any $\theta$.

The training proceeds as follows.
The output layer is initialized to zero, so that
$F_\theta \equiv 1$ at the start of optimization.
We then choose the weights in the loss function of Eq.~\eqref{eq:loss}
as $(w_E, w_\rho, w_\tau) = (10,0,0)$, $(10,1,0)$, or $(10,1,1)$ depending on the fitting strategy.
The dimensionless energy
residual
is
\begin{equation}
    r_E(\theta) = \frac{E(\theta) - E^\text{ref}}{1\, \text{Ha} \cdot N_\text{atoms}} - \theta_\text{Si},
\end{equation}
where $N_\text{atoms}$ is the number of atoms in the unit cell
and $\theta_\text{Si}$ is an additional learnable parameter that absorbs any per-atom energy offset,
leaving however all energy differences unchanged.
We then minimize the loss $\mathcal L$ jointly over all $34$ parameters
using a full-batch Broyden--Fletcher--Goldfarb--Shanno
algorithm with a backtracking line search,
as implemented in \textsc{Optim.jl}~\cite{mogensen2018optim},
with a limit of 40 iterations.

\section{Results}
\label{sec:results}

\subsection{Elastic properties of diamond}
\label{subsec:elastic_diamond}

\begin{figure}
    \centering
    \includegraphics[width=8.5cm]{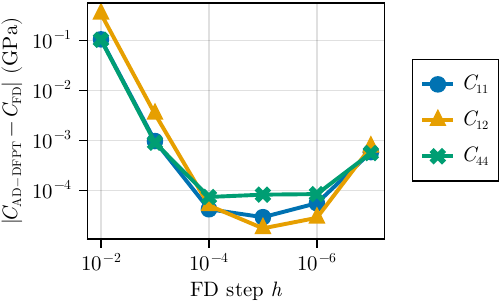}
    \caption{Comparison of AD-DFPT and finite difference (FD) clamped-ion elastic constants of diamond,
    computed for a range of FD step sizes $h$, with the r2SCAN01 XC functional.
    An excellent $<10^{-4}$~GPa agreement is observed, see also Table~\ref{tab:elastic_constants}.
    }
    \label{fig:elastic_findiff_convergence}
\end{figure}
\begin{table}
    \centering
    \begin{tabular}{l@{\hskip 0.5cm}l@{\hskip 0.3cm}l@{\hskip 0.3cm}l}
        \toprule
        & $C_{11}$ (GPa) & $C_{12}$ (GPa) & $C_{44}$ (GPa) \\
        \midrule
        AD-DFPT & \textbf{1110.6091}1 & \textbf{114.4044}6 & \textbf{584.554}32 \\
        FD $h = 10^{-5}$ & \textbf{1110.6091}4 & \textbf{114.4044}8 & \textbf{584.554}24 \\
        Prior work~\cite{Haxhijaj2026} & 1115 & 119 & 589 \\
        \bottomrule
    \end{tabular}
    \caption{Clamped-ion elastic constants of diamond, from AD-DFPT and centered finite differences (FD) of step $h=10^{-5}$
    computed using the r2SCAN01 XC functional (first two rows).
    For comparison, the last row displays previous relaxed-ion r2SCAN results~\cite{Haxhijaj2026}.}
    \label{tab:elastic_constants}
\end{table}

To validate our AD-DFPT implementation 
we compute the clamped-ion elastic constants
of diamond using both AD-DFPT and with finite differentiation (FD).
For our tests we employ the r2SCAN01 XC functional~\cite{Holzwarth2022r2scan01},
for which we have a norm-conserving pseudopotential generator at hand, see Sec.~\ref{sec:impl}.
Owing to its cubic structure, diamond has three independent elastic constants: $C_{11}$, $C_{12}$, and $C_{44}$.
Figure~\ref{fig:elastic_findiff_convergence} shows the difference between the AD-DFPT
and FD elastic constants for a range of step sizes $h$.
The smallest difference is reached at $h=10^{-5}$, and the corresponding elastic constants are listed
in Tab.~\ref{tab:elastic_constants}.
Our AD-DFPT elastic constants agree very well with FD, to within \SI{10^{-4}}{\giga\pascal}.
Furthermore, the agreement with the r2SCAN FD computations of Ref.~\onlinecite{Haxhijaj2026}
is within \SI{5}{\giga\pascal}.
This validates both our elastic-constant implementation and the AD-DFPT approach.
Note that in these tests the SCF has been converged very tightly to $10^{-12}$ in the density norm,
to limit FD numerical noise amplification.
Upon loosening the SCF tolerance, we observe that the AD-DFPT results remain more precise
than the FD ones, in agreement with our prior work~\cite{ADpaper}.

\subsection{Response properties of ZnO and \batio3}
\label{subsec:response_properties}
As a second application, we consider the computation of 
Born effective charges, relaxed-ion elastic constants, relaxed-ion piezoelectric tensor
and high-frequency and static dielectric tensor using a variety of DFT functionals:
the LDA with the Perdew--Wang parametrization~\cite{perdew1992lda},
the PBE GGA~\cite{Perdew1996},
and the r2SCAN01 meta-GGA.
Note that zero-point vibrational corrections are ignored throughout this work.

We compute these quantities following the procedure of \citeauthor{wu2005systematic}~\cite{wu2005systematic}.
First we obtain the elementary tensors provided in Table~\ref{tab:response_properties},
which can be expressed as derivatives $\tdfrac{A}{\theta}$ and computed using either plain AD-DFPT
or the modifications due to the Berry phase theory of polarization outlined in Sec.~\ref{subsec:elfield}.
From these we compute the relaxed-ion elastic tensor $C_{jk}$, relaxed-ion piezoelectric tensor $e_{\alpha j}$,
and relaxed-ion dielectric susceptibility $\chi_{\alpha\beta}$ (at fixed strain)
following Ref.~\onlinecite[Eqs.~(14)--(16)]{wu2005systematic}:
\begin{align}
    C_{jk} &= \bar{C}_{jk} - \Omega^{-1} \sum_{s\alpha t\beta} \Lambda_{s\alpha j} (K^{-1})_{s\alpha,t\beta} \Lambda_{t\beta k}, \\
    \label{eq:relaxed_piezoelectric_tensor}
    e_{\alpha j} &= \bar{e}_{\alpha j} + \Omega^{-1} \sum_{s\beta t\gamma} Z_{s\beta\alpha}^\ast (K^{-1})_{s\beta,t\gamma} \Lambda_{t\gamma j}, \\
    \chi_{\alpha\beta} &= \bar{\chi}_{\alpha\beta} + \Omega^{-1} \sum_{s\gamma t\delta} Z_{s\gamma \alpha}^\ast (K^{-1})_{s\gamma,t\delta} Z_{t\delta \beta}^\ast,
\end{align}
with $K^{-1}$ the pseudoinverse of the force constant matrix $K$.
Note that the Born effective charges are based on the \textit{total} polarization,
for which the ionic contribution $P_\alpha^\text{ion} = \frac{1}{\Omega} \sum_s Z_s R_{s\alpha}$
with ionic charges $Z_s$ and positions $\mathbf{R}_s$ is added to the
Berry-phase electronic polarization. %

Finally, we obtain the high-frequency dielectric tensor
\begin{equation}
    \epsilon^\infty_{\alpha\beta} = \delta_{\alpha\beta} + 4\pi \bar{\chi}_{\alpha\beta}
\end{equation}
and the static dielectric tensor
\begin{align}
    \epsilon^0_{\alpha\beta} = \delta_{\alpha\beta} + 4\pi \chi_{\alpha\beta}.
\end{align}
We study these quantities on two paradigmatic polar materials, wurtzite ZnO and rhombohedral \batio3.

\subsubsection{ZnO}
\begin{table*}
    \centering
    \begin{tabular}{l@{\hskip 1cm}r@{\hskip 0.6cm}r@{\hskip 0.2cm}rr}
        \toprule
        & \multicolumn{3}{c}{This work} & {\hskip 1cm}Experiment \\
        \cmidrule(r){2-4}
        Property & LDA & PBE & r2SCAN01 \\
        \midrule
        Structural parameters \\
        $a$ (\AA) & 3.184 & 3.276 & 3.232 & 3.247\footnote{\label{fn:zno_a}Reference~\onlinecite{yoshio2001zno}} \\
        $c$ (\AA) & 5.146 & 5.286 & 5.208 & 5.203\footnoteref{fn:zno_a} \\
        $u$ (dimensionless) & 0.379 & 0.379 & 0.379 & 0.381\footnoteref{fn:zno_a} \\
        \midrule
        Relaxed-ion elastic constants (GPa) \\
        $C_{11}$ & 222 & 189 & 209 & 209\footnote{\label{fn:zno_elastic}Reference~\onlinecite{LandoltBornstein_III29a}} \\
        $C_{12}$ & 140 & 108 & 127 & 120\footnoteref{fn:zno_elastic} \\
        $C_{13}$ & 124 & 93 & 115 & 104\footnoteref{fn:zno_elastic} \\
        $C_{33}$ & 239 & 201 & 216 & 216\footnoteref{fn:zno_elastic} \\
        $C_{44}$ & 38 & 37 & 39 & 44\footnoteref{fn:zno_elastic} \\
        \midrule
        Born effective charges (units of $e$) \\
        $Z^\ast_{\text{Zn},xx}$ & 2.085 & 2.136 & 2.094 & 2.02\footnote{Reference~\onlinecite{decremps2002zno}} \\
        $Z^\ast_{\text{Zn},zz}$ & 2.140 & 2.189 & 2.144 & \\
        \midrule
        Clamped-ion piezoelectric tensor (C/m$^2$) \\
        $\bar{e}_{31}$ & 0.35 & 0.37 & 0.35 & \\
        $\bar{e}_{33}$ & -0.72 & -0.74 & -0.70 & \\
        $\bar{e}_{15}$ & 0.38 & 0.39 & 0.37 & \\
        \midrule
        Relaxed-ion piezoelectric tensor (C/m$^2$) \\
        $e_{31}$ & -0.69 & -0.56 & -0.70 & -0.62\footnote{\label{fn:zno_piezo}Reference~\onlinecite{LandoltBornstein_III29b}} \\
        $e_{33}$ & 1.32 & 1.08 & 1.38 & 0.96\footnoteref{fn:zno_piezo} \\
        $e_{15}$ & -0.54 & -0.42 & -0.49 & -0.37\footnoteref{fn:zno_piezo} \\
        \midrule
        High-frequency dielectric tensor (units of $\epsilon_0$) \\
        $\epsilon^\infty_{11}$ & 5.10 & 5.27 & 4.30 & 3.70\footnote{\label{fn:zno_dielectric}Reference~\onlinecite{ashkenov2003zno_dielectric}} \\
        $\epsilon^\infty_{33}$ & 5.09 & 5.26 & 4.35 & 3.78\footnoteref{fn:zno_dielectric} \\
        \midrule
        Static dielectric tensor (units of $\epsilon_0$) \\
        $\epsilon^0_{11}$ & 9.43 & 10.56 & 8.82 & 7.77\footnoteref{fn:zno_dielectric} \\
        $\epsilon^0_{33}$ & 10.10 & 11.51 & 10.01 & 8.91\footnoteref{fn:zno_dielectric} \\
        \bottomrule
    \end{tabular}
    \caption{Structural and response properties of wurtzite ZnO obtained from AD-DFPT.}
    \label{tab:zno_table}
\end{table*}

ZnO is an attractive material for applications in electronics~\cite{ozgur2010zno_review},
among which piezoelectric devices,
and its reponse properties are known from numerous experimental measurements.
Table~\ref{tab:zno_table} lists our results for ZnO, listing both
the three independent structural parameters after relaxation
as well as the independent components of its response properties.

We find that r2SCAN01 predicts considerably more accurate lattice constants $a$ and $c$,
which only deviate by \SI{0.5}{\percent} and \SI{0.1}{\percent} from the experimental values, respectively.
In contrast the LDA and PBE results show larger deviations of $1$ to \SI{2}{\percent}.
It is thus not surprising that r2SCAN01 also predicts the best elastic constants,
with a mean absolute error (MAE) across the reported values of
\SI{16.4}{\giga\pascal} for LDA,
\SI{13.0}{\giga\pascal} for PBE,
and \SI{4.6}{\giga\pascal} for r2SCAN01.
For the Born effective charges all results are close to the nominal ionic charge of $2$.
In comparison with experiment~(only $Z^\ast_{\text{Zn},xx}$ available),
LDA and r2SCAN01 are of similar accuracy~(error of about \SI{3}{\percent})
while PBE deviates by \SI{5.7}{\percent}.

Considering the piezoelectric tensor,
PBE gives the best agreement with experiment.
Indeed, we find a MAE of
\SI{0.20}{\coulomb\per\m\squared} for LDA,
\SI{0.08}{\coulomb\per\m\squared} for PBE,
and \SI{0.21}{\coulomb\per\m\squared} for r2SCAN01.
Notably, all three functionals predict similarly small clamped-ion piezoelectric tensors~(see Table~\ref{tab:zno_table}),
so that the differences stem primarily from the ionic relaxation contribution
in Eq.~\eqref{eq:relaxed_piezoelectric_tensor},
which is generally of opposite sign to the clamped-ion tensor.
PBE generally underbinds in solids, which is reflected in a smaller ionic contribution
and hence, overall, in a piezoelectric tensor that is fortuitously closer to experiment.

Finally, we find that r2SCAN01 partially cures the overestimation of the
dielectric tensor of LDA and PBE.
For the high-frequency dielectric tensor $\epsilon^\infty$,
the MAE is $1.36$ for LDA,
$1.52$ for PBE,
and $0.58$ for r2SCAN01.
For the static dielectric tensor $\epsilon^0$,
the MAE is $1.42$ for LDA,
$2.70$ for PBE,
and $1.08$ for r2SCAN01.
This is not surprising: the overestimation of the dielectric tensor
is linked to the well-known underestimation of the band gap by semilocal XC functionals.
SCAN-family functionals are known to moderately improve the band gap over
over LDA and PBE~\cite{yang2016gks,isaacs2018scanperf}, thus also improving the obtained dielectric constants.

\subsubsection{\batio3}

\begin{table*}
    \centering
    \begin{tabular}{l@{\hskip 0.5cm}r@{\hskip 0.6cm}r@{\hskip 0.2cm}r@{\hskip 0.3cm}rrr}
        \toprule
        & \multicolumn{3}{c}{This work} \\
        \cmidrule(r){2-4}
        Property & LDA & PBE & r2SCAN01 & r2SCAN~\cite{Wang2026} & PBE0~\cite{mahmoud2014batio3} & Experiment~\cite{kwei1993batio3} \\
        \midrule
        Structural parameters \\
        $a$ (\AA) & 3.948 & 4.063 & 4.019 & 4.036 & 4.010 & 4.004 \\
        $\alpha$ (\unit{\degree}) & 89.93 & 89.69 & 89.80 & 89.82 & 89.80 & 89.84 \\
        $\Delta x_\text{Ti}$ (dimensionless) & 0.009 & 0.015 & 0.013 & 0.014\footnote{\label{fn:signflip}Sign flipped to match our spontaneous polarization direction} & 0.013\footnoteref{fn:signflip} & 0.013\footnoteref{fn:signflip} \\
        $\Delta x_\text{O}$ (dimensionless) & -0.009 & -0.016 & -0.013 & -0.011\footnoteref{fn:signflip} & -0.012\footnoteref{fn:signflip}\footnote{\label{fn:swapped}$\Delta x_\text{O}$ and $\Delta z_\text{O}$ exchanged compared to Ref.~\onlinecite{mahmoud2014batio3}} & -0.011\footnoteref{fn:signflip} \\
        $\Delta z_\text{O}$ (dimensionless) & -0.013 & -0.027 & -0.022 & -0.021\footnoteref{fn:signflip} & -0.023\footnoteref{fn:signflip}\footnoteref{fn:swapped} & -0.019\footnoteref{fn:signflip} \\
        \midrule
        Relaxed-ion elastic constants (GPa) \\
        $C_{11}$ & 300 & 224 & 262 & & 282 & \\
        $C_{12}$ & 94 & 61 & 73 & & 73 & \\
        $C_{13}$ & 57 & 23 & 33 & & 31 & \\
        $C_{14}$ & -44 & -42 & -46 & & -47\footnoteref{fn:signflip} & \\
        $C_{33}$ & 293 & 197 & 242 & & 258 & \\
        $C_{44}$ & 54 & 31 & 41 & & 44 & \\
        $C_{66}$ & 103 & 82 & 95 & & 104 & \\
        \midrule
        Born effective charges (units of $e$) \\
        $Z^\ast_{\text{Ba},xx}$ & 2.785 & 2.806 & 2.786 & 2.75 & & \\
        $Z^\ast_{\text{Ba},zz}$ & 2.759 & 2.719 & 2.727 & 2.70 & & \\
        $Z^\ast_{\text{Ti},xx}$ & 6.831 & 6.163 & 6.333 & 6.43 & & \\
        $Z^\ast_{\text{Ti},zz}$ & 6.241 & 4.805 & 5.282 & 5.41 & & \\
        \midrule
        Clamped-ion piezoelectric tensor (C/m$^2$) \\
        $\bar{e}_{21}$ & -0.20 & -0.32 & -0.28 & & -0.28 & \\
        $\bar{e}_{31}$ & -0.05 & -0.06 & -0.06 & & -0.06\footnoteref{fn:signflip} & \\
        $\bar{e}_{33}$ & 0.15 & 0.26 & 0.22 & & 0.22\footnoteref{fn:signflip} & \\
        $\bar{e}_{24}$ & -0.16 & -0.21 & -0.20 & & -0.21\footnoteref{fn:signflip} & \\
        \midrule
        Relaxed-ion piezoelectric tensor (C/m$^2$) \\
        $e_{21}$ & 3.89 & 1.64 & 2.16 & & 1.99 & \\
        $e_{31}$ & 3.58 & 1.90 & 2.29 & & 2.17\footnoteref{fn:signflip} & \\
        $e_{33}$ & 5.07 & 3.19 & 3.58 & & 3.45\footnoteref{fn:signflip} & \\
        $e_{24}$ & 7.01 & 3.62 & 4.40 & & 4.67\footnoteref{fn:signflip} & \\
        \midrule
        High-frequency dielectric tensor (units of $\epsilon_0$) \\
        $\epsilon^\infty_{11}$ & 6.28 & 5.68 & 5.39 & 5.53 & & \\
        $\epsilon^\infty_{33}$ & 5.99 & 5.03 & 4.92 & 5.05 & & \\
        \midrule
        Static dielectric tensor (units of $\epsilon_0$) \\
        $\epsilon^0_{11}$ & 104.11 & 39.03 & 48.41 & & & \\
        $\epsilon^0_{33}$ & 53.07 & 21.07 & 24.62 & & & \\
        \bottomrule
    \end{tabular}
    \caption{Structural and response properties of rhombohedral \batio3 obtained from AD-DFPT.}
    \label{tab:batio3_table}
\end{table*}

Next we consider \batio3, a prototypical ferroelectric material.
At high temperatures, it has a cubic perovskite structure,
but as the temperature is lowered
it undergoes a series of phase transitions that lower its symmetry
to a tetragonal, then orthorhombic and then finally rhombohedral structure.
We study the response properties of its low-temperature rhombohedral phase,
which are difficult to measure experimentally.
We thus compare our predictions with previous results%
~\cite{mahmoud2014batio3,Wang2026}
based on the PBE0 hybrid functional~\cite{adamo1999pbe0,ernzerhof1999pbe0}
as well as the r2SCAN functional.
Note that the occasionally quoted experimental dielectric constants
of $\epsilon^\infty_{11} = 6.19$ and $\epsilon^\infty_{33} = 5.88$
from Ref.~\onlinecite{Wang2001batio3} in fact refer to the
room-temperature tetragonal phase of \batio3~\cite{mahmoud2014batio3}.

The cell parameters of \batio3 are the lattice constant $a$
and the rhombohedral angle $\alpha$.
For the ionic positions,
we use the parametrization of Ref.~\onlinecite{kwei1993batio3},
for which
Ba is at $(0, 0, 0)$,
Ti at $(0.5 + \Delta x_\text{Ti}, 0.5 + \Delta x_\text{Ti}, 0.5 + \Delta x_\text{Ti})$,
and the first O is at $(0.5 + \Delta x_\text{O}, 0.5 + \Delta x_\text{O}, \Delta z_\text{O})$,
all in relative coordinates.
Note that we use the opposite spontaneous polarization convention compared to previous works~\cite{Wang2026,mahmoud2014batio3,kwei1993batio3}.
Consequently, the ionic displacements $\Delta x_\text{Ti}$,
$\Delta x_\text{O}$, and $\Delta z_\text{O}$,
as well as the tensor entries with an odd number of indices referring to Cartesian direction $3$ ($z$)
or Voigt component $4$ ($yz$) or $5$ ($xz$),
are of the opposite sign.
We adapted all values displayed in Table~\ref{tab:batio3_table}
to our sign convention.

Without surprise we find that the structural parameters and elastic
constants of r2SCAN01 are in better agreement with experiment
and hybrid DFT than the values resulting from LDA and PBE.
For the Born effective charges and dielectric constants
only limited reference data is available. However, we find
a very good agreement between our r2SCAN01 results
and previous r2SCAN computations~\cite{Wang2026}.

Finally, compared to the other functionals considered here our r2SCAN01 calculations
predict relaxed-ion piezoelectric tensors
in best agreement with the PBE0 reference:
the MAE is \SI{1.82}{\coulomb\per\m\squared} for LDA,
\SI{0.48}{\coulomb\per\m\squared} for PBE,
and \SI{0.17}{\coulomb\per\m\squared} for r2SCAN01.
In \batio3 the piezoelectric tensor is dominated
by a soft mode, such that the ionic contributions
are an order of magnitude larger than the
clamped-ion piezoelectric tensors.
While r2SCAN01 captures the ionic contributions well%
---on top of a near-perfect agreement with PBE0 for the clamped-ion contributions---%
these are largely overestimated by LDA
and underestimated by PBE.

Overall, our results demonstrate
our AD-DFPT implementation to be sufficiently
versatile to access a wide range of response properties.
Importantly, being able to seamlessly consider meta-GGA level functionals
(here r2SCAN01) is promising and brings predictions noticeably
closer to experimental or hybrid-DFT reference values. 
\subsection{
Density-based XC functional optimization}
\label{subsec:xc_learning}

As a final application we illustrate the gradient-based learning
of XC functional parameters $\theta$
by minimizing a loss function $\mathcal L$ that includes the self-consistent density. 
Based on AD-DFPT the required derivatives $\tdfrac{\rho}{\theta}$,
or even directly $\tdfrac{\mathcal L}{\theta}$,
are obtained accurately and conveniently,
 see Sec.~\ref{subsec:ad_dfpt}.

More precisely, we show the training of
a meta-GGA exchange functional for silicon to reproduce costly hybrid DFT results.
For each structure, we performed a calculation with the HSE06~\cite{hse03,hse06} functional
yielding reference total energy, electronic density,
and KED computed from the hybrid orbitals.
For the standard diamond phase of silicon and its high-pressure $\beta$-tin phase,
reference computations were performed at seven different volumes.
Only data from three of the diamond-phase structures are included in the training set,
and we compare three different strategies to fit our exchange functional to this data:
(i) only making use of the total energies,
(ii) using total energies and the densities $\rho$,
as well as
(iii) using the total energies, densities $\rho$, and the KED $\tau$.
This is achieved through the loss function
\begin{widetext}
\begin{align}
\mathcal{L}(\theta) =
\mathbb{E}_{\text{train}}
\Bigg[
 w_E \underbrace{r_E(\theta)^2}_{\text{energy loss}}
+w_\rho \underbrace{\frac{\|\rho(\theta) - \rho^\text{ref}\|^2}{\|\rho^\text{ref}\|^2}}_{\text{density loss}}
+w_\tau \underbrace{\frac{\|\tau(\theta) - \tau^\text{ref}\|^2}{\|\tau^\text{ref}\|^2}}_{\text{KED loss}}
\Bigg],
\label{eq:loss}
\end{align}
\end{widetext}
with varying weights $w_E$, $w_\rho$, $w_\tau$, 
and $\|.\|$ is the $L^2$ norm.
See Sec.~\ref{subsec:mlxcdetails} for the used weights and
the definition of the dimensionless
energy residual
term $r_E(\theta)$.

As the ansatz for our XC functional we consider
a fixed PBE correlation $E_{\mathrm{c}}^\mathrm{PBE}$
(same as the HSE06 reference) as well as an exchange part built
from a learned meta-GGA semilocal exchange enhancement factor $F_\theta$
over the LDA exchange energy density $e_\mathrm{x}^\mathrm{LDA}$:
\begin{align}
    E_\mathrm{xc}(\rho,\tau; \theta) &= E_\mathrm{x}(\rho,\tau;\theta) + E_{\mathrm{c}}^\mathrm{PBE}(\rho), \\
    \label{eq:ml_x_functional}
    E_\mathrm{x}(\rho, \tau; \theta) &= \int_\Omega\dif \mathbf r\,
    e_\mathrm{x}^\mathrm{LDA}(\rho(\mathbf r))\, F_\theta(\rho(\mathbf r), |\nabla\rho(\mathbf r)|, \tau(\mathbf r)).
\end{align}
In our model $F_\theta$ is a $33$-parameter neural network taking two bounded
dimensionless variables that vanish for the homogeneous electron gas,
see Sec.~\ref{subsec:mlxcdetails} for more details.
The output layer is zero-initialized, so that $F_\theta \equiv 1$ and optimization begins from
LDA exchange with PBE correlation.
While the energy-only fit (i) here is strongly underdetermined,
the inclusion of the density (ii) and KED (iii) provides spatially 
resolved information as additional constraints.

\begin{figure*}
    \includegraphics[width=\textwidth]{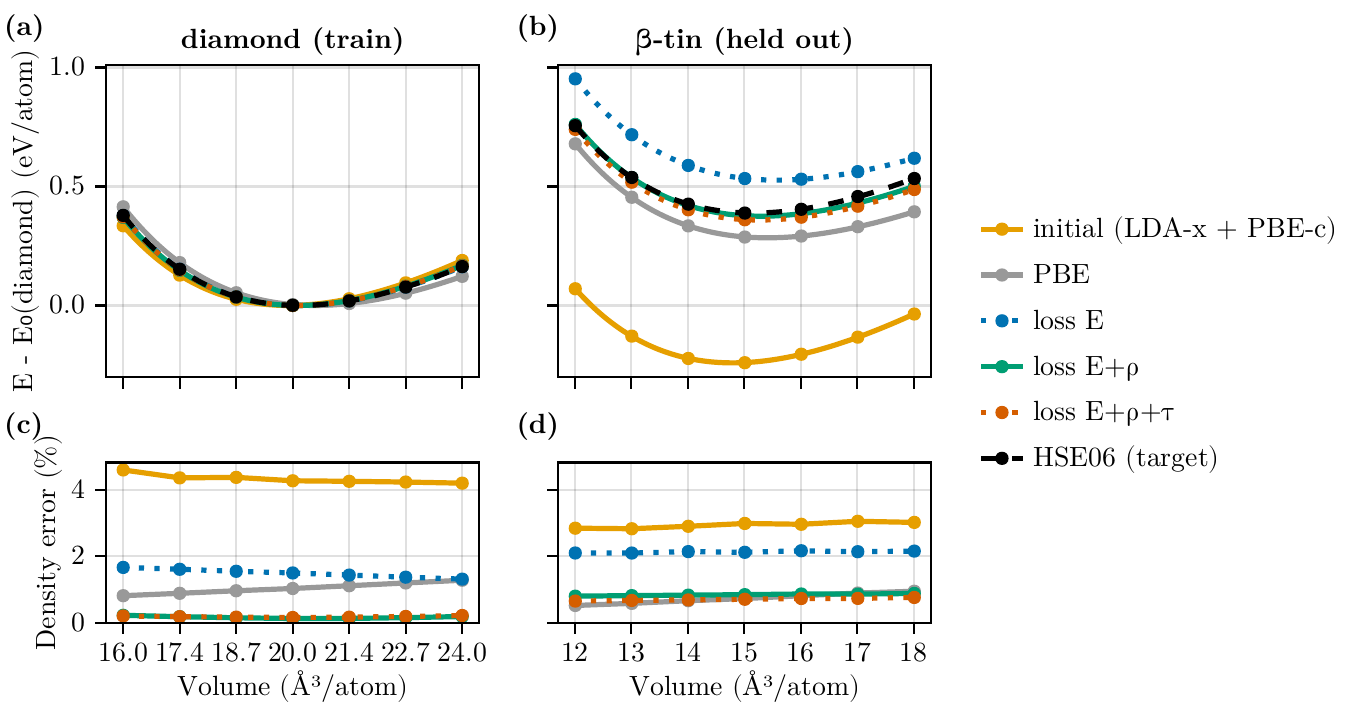}
    \caption{Transferability of meta-GGA functionals trained on three volumes of
    diamond-phase silicon under the loss variants of Eq.~\eqref{eq:loss}, compared to: the initial
    functional the training starts from (LDA exchange with PBE correlation),
    PBE, HSE06 target. (a)(b)~Energy-volume curves for the trained
    diamond phase and the held-out $\beta$-tin phase, each functional aligned to its own
    fitted diamond minimum, so that the inter-phase energy difference is comparable across methods. (c)(d)~Corresponding
    self-consistent density
    $L^2$-errors
    relative to the HSE06 reference.
    }
    \label{fig:lossvariants}
\end{figure*}

The top of
Fig.~\ref{fig:lossvariants} shows the obtained equations of state (EOS)
for the diamond and $\beta$-tin phases of silicon
computed with the XC functional
from each of the three training strategies.
For each structure of the EOS fit,
the bottom half of the figure displays
the relative $L^2$ density error against the HSE06 density.
For comparison, the HSE06 reference,
our starting point before training
(i.e.~LDA exchange plus PBE correlation)
as well as PBE predictions
are also shown.%

The starting functional is a poor
description of silicon: 
comparing the equilibrium energies for both phases,
it energetically favors the $\beta$-tin phase
over the correct diamond phase by about~\SI{0.2}{\eV/atom}.
Moreover, its self-consistent density deviates from the
HSE06 reference by several percent throughout.
On the diamond phase (Fig.~\ref{fig:lossvariants}, left column),
the density error is large for
the energy-only training strategy (i),
but the trained functional nonetheless manages to almost perfectly
reproduce the diamond silicon EOS.
Including $\rho$ and $\tau$~(strategies (ii) and (iii))
only marginally affects the diamond silicon EOS.

In contrast, when considering the transferability
of the fitted functionals to the unseen $\beta$-tin phase of silicon (Fig.~\ref{fig:lossvariants}, right column),
notable deviations between the training strategies can be observed.
For energy-only training both the EOS as well as
the predicted densities do deviate significantly from the HSE06 reference.
However, the transferability of training strategies (ii) and (iii) is improved:
the EOS is very close to the HSE06 reference
even though no $\beta$-tin structure was included in the training set.

Notably, our functionals trained on $\rho$-data do
outperform PBE in the EOS predictions even though the PBE densities
agree similarly well with HSE06 as our $\rho$-trained models.
Furthermore, we observe no significant difference between training strategies (ii) and (iii),
suggesting that the added KED loss makes little difference here;
fitting only on $\rho$ could have been sufficient.
Further tests are needed to confirm the generality of these observations.
Overall, this toy example illustrates how training against reference densities
is feasible thanks to AD-DFPT and can lead to better data efficiency and transferability.

\section{Conclusion}
\label{sec:conclusion}
We presented a plane-wave formulation of DFPT
for XC functionals with a dependence on the kinetic energy density~(KED).
Our framework covers
any functional of the general form $E_\mathrm{xc}(\rho, \tau)$,
which includes, but is not limited to, semilocal meta-GGAs.
To achieve such generality,
we reformulated the computation
of the required second derivatives of the XC energy within the DFPT procedure
as a Jacobian-vector product~(JVP),
which lends itself to efficient evaluation
using algorithmic differentiation~(AD) techniques.
By integrating this extended core DFPT solver into our AD-DFPT framework~\cite{ADpaper},
the computation
of the derivative of any ground state quantity of interest wrt.~any perturbation
for any such functional $E_\mathrm{xc}(\rho, \tau)$ is now feasible.
Beyond standard energy derivatives~(e.g.~elastic constants) and
unusual derivatives (such as the change of the density wrt.~XC functional parameters),
we detailed the computation of polarization responses and electric field perturbations using AD-DFPT.
As a result, a sizeable set of key material properties is now available in the AD-DFPT formalism.

We demonstrated these capabilities in two showcases.
First, we computed a range of standard properties
for ZnO and \batio3 at the meta-GGA level
for which our results confirm that meta-GGA functionals (here r2SCAN01)
often noticeably outperform LDA and PBE.
Second, we considered the density-based learning of a neural-network meta-GGA functional
using gradients computed by AD-DFPT.
The
ability to include density information during training
improved the transferability of our machine-learned meta-GGA functional,
compared to fitting against reference energies only.

An obstacle to broader testing of such density-based XC learning strategies is
that our current AD-DFPT framework relies on forward-mode AD.
As a result, evaluating the required XC parameter gradients
requires one full DFPT calculation per XC parameter,
which limits its applicability to a modest number of parameters.
In contrast, reverse-mode AD techniques promise a computational scaling
that is independent of the number of XC parameters.
Developing such techniques is crucial
to scaling AD-DFPT-based XC training
and to enabling the training of XC functionals
based on modern neural network architectures.

We remark that consistent
plane-wave-based meta-GGA calculations %
are currently limited by the scarcity of
matching norm-conserving meta-GGA pseudopotentials.
However, a number of recent works are devoted to developing
generally applicable meta-GGA pseudopotentials~\cite{metapsp,gareis2026,Holzwarth2022r2scan01},
such that this gap may soon be closed.

Overall, our extension of plane-wave DFPT
ties fully into the opportunities enabled by differentiable DFT workflows~\cite{ADpaper},
essentially extending them to the broader class of KED-dependent functionals.
In this work we explicitly discussed XC learning,
but other exciting prospects for future work
include computing DFT uncertainties in meta-GGA predictions,
e.g.~by applying the ideas outlined in Ref.~\onlinecite{ADpaper}
to XC functionals with parameter uncertainty~\cite{wellendorff2014mbeef,lundgaard2016mbeefvdw,Aldegunde2016,brown2021,kai2022}.

\section*{Acknowledgement}
We would like to express our gratitude to Giovanni Pizzi
and Timo Reents for sharing an early version of their reference meta-GGA
all-electron equations of state.
We further thank Stefan Riemelmoser and Austin Zadoks
for stimulating discussions.
This research was supported by the Swiss National Science Foundation (SNSF, Grant No.~10002757) as well as the NCCR MARVEL, a National Centre of Competence in Research, funded by the SNSF (Grant No.~205602).

We used Claude Opus 4.6 to 5.5 from Anthropic
to support the implementation of our DFPT algorithm,
the pseudopotential verification in the SM,
and the numerical experiments in Sec.~\ref{subsec:xc_learning}.
All code was subsequently human-verified.
GLM-5.3-Flash from Z.ai was employed for lightweight text editing.
All authors cross-read
and approved the final version of the text.

\section*{Data Availability Statement}
The data and code supporting the findings of this study are openly available in a public repository on GitHub at \repourl,
archived at Ref.~\onlinecite{supporting_repo}.
The repository contains all datasets and implementation details necessary to reproduce the numerical results presented in this article.

\bibliographystyle{apsrev4-2}
\bibliography{refs}

\appendix
\cleardoublepage
\input{supplement.tex}

\end{document}

%% file: supplement.tex
% \theH... versions are needed for hyperref's links
\setcounter{section}{0}
\renewcommand{\thesection}{S\arabic{section}}
\renewcommand{\theHsection}{S\arabic{section}}
\setcounter{figure}{0}
\renewcommand{\thefigure}{S\arabic{figure}} % S1, S2, ...
\renewcommand{\theHfigure}{S\arabic{figure}}
\setcounter{table}{0}
\renewcommand{\thetable}{S\arabic{table}}
\renewcommand{\theHtable}{S\arabic{table}}
\setcounter{equation}{0}

\onecolumngrid  % Switch to one-column mode
\begin{center} \Large Supplementary Material \end{center}

\section{Accuracy of the r2SCAN01 pseudopotentials}
\label{supp:pseudos}
In this section, we compare our r2SCAN01 pseudopotentials
to the r2SCAN01 reference all-electron (AE) data by \citeauthor{mgga_ae_data}~\cite{mgga_ae_data}
The comparison is based on the verification protocol by Bosoni et al.~\cite{Bosoni2023}
For each element, and for 4 unary configurations---%
body-centered cubic (BCC), diamond, face-centered cubic (FCC), simple cubic (SC)---%
we evaluate the total energy within pseudopotential DFT for a range of unit cell volumes.
This data is fitted against a Birch--Murnaghan equation of state parametrization,
and the fit parameters are compared to the AE reference.
To compare the equations of state, we use the $\varepsilon$ and $\nu$ metrics~\cite{Bosoni2023}.
An excellent agreement is $\varepsilon < 0.06$ or $\nu < 0.1$,
while a good agreement is $\varepsilon < 0.2$ or $\nu < 0.33$.

Table~\ref{tab:pseudo_epsilon} lists
the $\varepsilon$ and $\nu$ metrics respectively
for our 5 pseudopotentials across the 4 unary configurations.
Our pseudopotentials are within the good agreement threshold,
save FCC Ba ($\varepsilon = 0.21$ and $\nu = 0.37$) and SC Zn ($\nu = 0.35$).
We thus consider our pseudopotentials to be of sufficient accuracy
for our testing purposes.

\begin{table}[h]
    \centering
    \begin{tabular}{lrrrr}
    \toprule
    \multicolumn{5}{c}{$\varepsilon$ metric}\\
    \midrule
    & BCC & Diamond & FCC & SC \\
    \midrule
    Ba & 0.01 & 0.09 & \textbf{0.21} & 0.12 \\
    C  & 0.05 & 0.06 & 0.12 & 0.19 \\
    O  & 0.04 & 0.04 & 0.02 & 0.07 \\
    Ti & 0.06 & 0.03 & 0.03 & 0.08 \\
    Zn & 0.04 & 0.06 & 0.12 & 0.17 \\
    \bottomrule
    \end{tabular}
    \hspace{3em}
    \begin{tabular}{lrrrr}
    \toprule
    \multicolumn{5}{c}{$\nu$ metric}\\
    \midrule
    & BCC & Diamond & FCC & SC \\
    \midrule
    Ba & 0.10 & 0.14 & \textbf{0.37} & 0.20 \\
    C  & 0.08 & 0.10 & 0.22 & 0.31 \\
    O  & 0.06 & 0.06 & 0.04 & 0.14 \\
    Ti & 0.09 & 0.18 & 0.05 & 0.13 \\
    Zn & 0.09 & 0.17 & 0.18 & \textbf{0.35} \\
    \bottomrule
    \end{tabular}
    \caption{$\varepsilon$ metric (left)
    and $\nu$ metric (right)
    comparison of our r2SCAN01 pseudopotentials with AE reference data.
    Cases failing to be in at least good agreement are marked in bold.}
    \label{tab:pseudo_epsilon}
\end{table}